# A (T-P) phase diagram of hydrogen storage on $(N_4C_3H)_6Li_6$ and the possibility of an associated new functional material

R. Das and P. K. Chattaraj*

Department of Chemistry and Center for Theoretical Studies, Indian Institute of Technology Kharagpur, Kharagpur – 721302, West Bengal, India

*Author for correspondence: Email: pkc@chem.iitkgp.ernet.in

**ABSTRACT**

A temperature – pressure phase diagram has been generated through the study of hydrogen adsorption on the $(N_4C_3H)_6Li_6$ cluster at the B3LYP/6-31G(d) level of theory. Possibility of hydrogen storage in an associated 3D functional material has been also explored.

1. INTRODUCTION

Hydrogen, being considered as sustainable energy carrier, has received increasing attention in the area of experimental and theoretical research since it is renewable and environment friendly. There are two ways to store hydrogen, either as liquid, only at a very low temperature or as gas, only at extremely high pressure[1]. Both these two ways have some drawbacks, storing hydrogen as liquid causes high energy expenditure to maintain proper cryogenic environment where as gas storage may cause explosion. Further, the use of hydrogen as energy source is facing a major problem due to lack of proper storage materials. An ideal hydrogen storage material should store hydrogen in high volumetric and gravimetric densities with fast kinetics as well as favorable thermodynamics[2]. Hence, developing new material for effective and safe hydrogen storage is quite a challenging task.

Considerable efforts have been devoted towards developing new storage materials. Various materials have been investigated such as metal hydrides[3,4], polymers[5-7], metal organic framework (MOF)[8-10], covalent organic framework (COF)[11], carbon-based nanomaterials[12-14], clathrates[15-18] etc. But most of these materials fail to meet all the requirements for commercial hydrogen storage set by the Department of Energy (DOE), USA[19]. Hence, to store hydrogen in high gravimetric wt%, materials consisting of light molecules such as carbon, boron are more appropriate. Therefore the materials like carbon nano-tubes (CNT)[20], boron nanomaterials[21], fullerenes[22], graphenes[23] are potential candidates for hydrogen storage. However, some scientific studies reveal that pure carbon materials interact weakly with hydrogen[24]. Later some research groups explored the idea that doping of carbon materials with light alkali metals[25] or substitution of some *C* centers by *B* atoms[26] improves the binding energy of hydrogen.

Recently, metal clusters like trigonal cationic $Li_3^+$, $Na_3^+$ rings[27] and neutral $Mg_n$ and $Ca_n$ [$n = 8 - 10$] cages[27] (analogues to metastable hydrogen – stacked $Mg_n$ clusters) have been checked for their ability to be used as a potential hydrogen (both in atomic and molecular form) trapping materials. Further, some recent studies have shown that $B_{80}$[28] and BN bucky balls[29] can be executed for trapping hydrogen in atomic or molecular form depending on their trapping mode (exohedral or endohedral trapping).

Again, several experimental and theoretical studies have shown that both temperature and pressure have some effects on hydrogen storage capacity of different materials like zeolite[1], ZnO sheet[30], MOF[31], clathrate hydrates[32] etc. For most of these materials the gravimetric densities of hydrogen storage reaches considerably high values only at a very low temperature and high pressure.

Chattaraj et al.[33] in their recent study have shown that metal bonded $N_4^{2-}$ and $N_6^{4-}$ can act as effective storage materials.

In this report, we have doped planar crown like $(N_4C_3H)_6$ cluster[34] reported earlier with Li atom and performed conceptual density functional theory (CDFT) based calculations to study the hydrogen adsorption on $(N_4C_3H)_6Li_6$ cluster. We have found that each Li center can adsorb three hydrogen molecules at temperature 298.15K and pressure 1atm. We have calculated ΔG values for hydrogen adsorption process on $(N_4C_3H)_6Li_6$, resulting $6H_2@(N_4C_3H)_6Li_6$ as product, at different temperature and pressure and have shown the variation of ΔG value through a temperature – pressure phase diagram.

Here we have also designed a 3D functional material[35], considering two planar $(N_4C_3H)_6$ moieties as building blocks which have been connected by –C≡C– linkers. We have also studied the hydrogen adsorption on this 3D structure and we found that the cluster can adsorb twelve hydrogen molecules.

2. **THEORETICAL BACKGROUND**

The stability of a molecular system in its ground state is characterized by its existence at the minimum on the potential energy surface (PES). This condition of stability can be complemented by various associated electronic structure principles of maximum hardness[36-38] ($\eta$) and minimum polarizability[39,40] ($\alpha$) and minimum electrophilicity[41,42] ($\omega$).

In the present work we have used conceptual density functional theory (CDFT) based global reactivity descriptors like electronegativity[43-45] ($\chi$), hardness[46-48] ($\eta$) and electrophilicity[49-51] ($\omega$) as tools for analyzing the stability and reactivity of the studied systems.

For an N-electron system having total energy E, the electronegativity[43-45] ($\chi$) and hardness[46-48] ($\eta$) can be defined as follow

$$\chi = - \left(\frac{\partial E}{\partial N}\right)_{v(r)} = - \mu \qquad (1)$$

$$\eta = \left(\frac{\partial^2 E}{\partial N^2}\right)_{v(r)} \qquad (2)$$

where $v(r)$ and $\mu$ are external and chemical potentials respectively. Electrophilicity[49-51] ($\omega$) is defined as

$$\omega = \frac{\mu^2}{2\eta} = \frac{\chi^2}{2\eta} \tag{3}$$

Applying the finite difference approximations to equations 1 and 2, $\chi$ and $\eta$ can be expressed as

$$\chi = \frac{I+A}{2} \tag{4}$$

and

$$\eta = I - A \tag{5}$$

where $I$ and $A$ are the ionization potential and electron affinity of a system respectively and are computed using the energies of the N-electron system with energy E(N). They may be expressed as follow:

$$I = E(N-1) - E(N) \tag{6}$$

$$A = E(N) - E(N+1) \tag{7}$$

## 3. COMPUTATIONAL DETAILS

The molecular geometries of $(N_4C_3H)_6Li_6$ cluster and its corresponding hydrogen trapped analogues $nH_2@(N_4C_3H)_6Li_6 (n = 6,12,18,24)$ (shown in Fig. 1) have been optimized and characterized at the B3LYP/6-31G(d) level of theory by using the GAUSSIAN 03[52] program suite. Harmonic vibrational frequency analysis has been also performed for all the structures at the same level of theory. The number of imaginary frequency (NIMAG) values for all the optimized geometries turn out to be zero confirming their existence at minima on the potential energy surface (PES). The single point calculations for all the structures except for $24H_2@(N_4C_3H)_6Li_6$ have been performed to evaluate the energies of (N ± 1) electron systems considering the geometries of the corresponding optimized N-electron systems. The $I$ and $A$ values have been calculated using the ΔSCF technique. The electronegativity ($\chi$), hardness ($\eta$) and electrophilicity ($\omega$) have been calculated using equations 4, 5 and 3 respectively. The calculations of atomic charges ($Q_k$) have been carried out adopting the natural population analysis (NPA) scheme. The nucleus independent chemical shift[53] NICS (0) values have been calculated at the centers of the $(N_4C_3H)_6Li_6$ ring and corresponding hydrogen bound systems, i.e., $nH_2@(N_4C_3H)_6Li_6 (n = 6,12,18)$. The structures of $(N_4C_3H)_6Li_6$, $6H_2@(N_4C_3H)_6Li_6$ and also molecular hydrogen have been optimized at different temperature and pressure at the same level of theory to study the effect of temperature and pressure on ΔG for hydrogen adsorption process. The 3D functional moiety and its hydrogen trapped structure have been also optimized at the B3LYP/6-31G(d) level of theory by using the GAUSSIAN 03 program suite and frequency calculation has been also performed. The number of imaginary frequency (NIMAG) value for the 3D structure turns out to be zero, confirming its existence at the minima on the potential energy surface. However, we have found an imaginary frequency -14.397 cm$^{-1}$, for the hydrogen trapped structure of the 3D functional material.

To check the feasibility of hydrogen trapping reactions on $(N_4C_3H)_6Li_6$ and the stability of the corresponding hydrogen bound clusters the average chemisorption energy ($E_{CE}$) and interaction energy per molecule ($E_{IN}$) have been calculated from the energies of hydrogen adsorbed clusters ($E_{X(H_2)_n}$), parent moiety ($E_X$) and hydrogen molecule ($E_{H_2}$) using the following equations

$$E_{CE} = [E_X + nE_{H_2} - E_{X(H_2)_n}]/n \tag{8}$$

$$E_{IN} = [E_{X(H_2)_n} - (E_X + nE_{H_2})]/n \tag{9}$$

where $n$ is the no of hydrogen molecules.

4. RESULTS AND DISCUSSION

Figure 1 shows the optimized geometry of $(N_4C_3H)_6$ cluster, its Li doped structure, i.e. $(N_4C_3H)_6Li_6$ and all the hydrogen adsorbed structures $nH_2@(N_4C_3H)_6Li_6$ ($n = 6,12,18,24$). For all these structures number of imaginary frequencies turns out to be zero, confirming their existence at minima on the potential energy surface (PES) at the used level of theory. It is clear from this figure that three hydrogen molecules remain bound to the Li center whereas the fourth hydrogen molecule moving away from the Li center. Hence, each Li center can adsorb up to three hydrogen molecules as a result the $(N_4C_3H)_6Li_6$ cluster can adsorb 18 hydrogen molecules.

The point group (PG), total energy (E,au), and global reactivity parameters like electronegativity ($\chi$, $eV$), hardness ($\eta$, $eV$) and electrophilicity ($\omega$, $eV$) values for $(N_4C_3H)_6Li_6$ cluster and its corresponding hydrogen trapped analogues, i.e., $nH_2@(N_4C_3H)_6Li_6$ ($n = 6,12,18$) have been summarized in Table 1. It is clear that the energy decreases with gradual increase in the size of the clusters. The electronegativity ($\chi$) values are positive for $(N_4C_3H)_6Li_6$ and its hydrogen bound analogues, this indicates that all these system can take electrons upon chemical response. Table 1 also reveals that global hardness ($\eta$) increases in the following order $\eta_{(N_4C_3H)_6Li_6} < \eta_{6H_2@(N_4C_3H)_6Li_6} < \eta_{12H_2@(N_4C_3H)_6Li_6} < \eta_{18H_2@(N_4C_3H)_6Li_6}$. Hence, it is clear that upon hydrogen loading the resulting complex becomes harder than the parent moiety, i.e., $(N_4C_3H)_6Li_6$ cluster (shown in Figure 2a). This trend in hardness ($\eta$) has been well anticipated by a gradual decrease in electrophilicity ($\omega$) values (shown in Figure 2b). Therefore the clusters get stabilized on hydrogen adsorption vis-à-vis the validity of the principles of maximum hardness[36-38] and minimum electrophilicity[41,42].

The interaction energy per molecule ($E_{IE}$, Kcal/mol), average chemisorption energy ($E_{CE}$, Kcal/mol) and reaction enthalpy ($\Delta H$, Kcal/mol) values as calculated by considering gradual trapping of hydrogen molecule by $(N_4C_3H)_6Li_6$, have been reported in Table 2. The interaction energy of the reaction involving Li atom and $(N_4C_3H)_6$ is quite negative which ensures viability of the reaction. Table 2 reveals that the interaction energy calculated for hydrogen adsorption processes are negative for all the cases. As the number of the hydrogen molecule to be adsorbed increases the interaction between Li center and hydrogen molecule decreases (as the interaction energy becomes less negative). The average chemisorption energy given in Table 2 are all positive which indicates a favorable hydrogen adsorption on the metal surface. However, the value of average chemisorption energy ($E_{CE}$) diminishes with increasing number of hydrogen molecule to be adsorbed. This indicates that with increase in the number of hydrogen molecules around the metal center the adsorption process becomes less feasible due to possible steric crowding. It is also important to note that a good hydrogen storage material should allow the adsorption which is stronger than physisorption but weaker than chemisorption. The desorption behavior is also important. The favorable hydrogen adsorption by $(N_4C_3H)_6Li_6$ may be further established by the associated reaction enthalpy value. The negative reaction enthalpy values reported in Table 2 suggest that all the hydrogen trapping procedures are exothermic in nature, thereby justifying the stability of the resulting hydrogen adsorbed systems.

The charges on the Li centers of $(N_4C_3H)_6Li_6$ and its corresponding hydrogen bound clusters, i.e., $nH_2@(N_4C_3H)_6Li_6$ ( $n = 6,12,18$) clusters and their NICS(0) values have been tabulated in Table 3. A detailed analysis of Table 3 shows that all the Li centers of $(N_4C_3H)_6Li_6$ bear a positive charge which reflects a considerable charge transfer from Li to $(N_4C_3H)_6$ cluster. This positively charged Li center induces a dipole moment in the approaching hydrogen molecule, this phenomenon helps the binding of the hydrogen through a charge-induced dipole interaction. As a result one hydrogen atom becomes partially positively charged and the other one becomes partially negatively charged. The charge on Li center decreases as the number of hydrogen molecules around Li center increases (shown in Figure 2d), which leads to a decreasing Li-H interaction (illustrated in Figure 2c). The H-H bond length in the adsorbed hydrogen molecule is longer compared to that in the free hydrogen molecule (optimized at B3LYP/6-31G(d) level).

The NICS (0) values for $(N_4C_3H)_6Li_6$ and $nH_2@(N_4C_3H)_6Li_6$ ($n = 6,12,18$) clusters are slightly positive which indicates that $(N_4C_3H)_6Li_6$ and its hydrogen adsorbed analogues are nonaromatic in nature. The crown cluster, i.e., $(N_4C_3H)_6$ is an aromatic system[34] (NICS(0)= -7.480 ppm) which becomes nonaromatic on Li doping.

The variation of ΔG value for the hydrogen adsorption process with temperature (T) and pressure (P) has been presented in a T-P phase diagram (Figure 3) considering hydrogen adsorption on $(N_4C_3H)_6Li_6$ resulting in $6H_2@(N_4C_3H)_6Li_6$ as the hydrogen bound system. In this diagram the ΔG values below the black line are all negative. Therefore in this temperature-pressure region (ΔG<0) hydrogen adsorption process is thermodynamically favorable and the resulting hydrogen bound system is stable. The ΔG values above the red line are all positive which indicates that in this temperature-pressure region (ΔG>0) backward reaction may be favorable resulting in dissociation of hydrogen molecule from the cluster – hydrogen system. The region sandwiched by these two lines may have ΔG = 0 value which corresponds to the equilibrium state.

The optimized structures of the 3D molecular material and its hydrogen trapped analogue have been presented in Figure 4. We have found an imaginary frequency -14.397 cm$^{-1}$, for the hydrogen trapped structure of the 3D functional material. Hence, the structure is approximately very close to one of the minima of on the PES. This 3D moiety can adsorb up to12 hydrogen molecules. Further work is in progress.

## 5. CONCLUSION

The DFT based calculation reveals that the $(N_4C_3H)_6Li_6$ cluster and corresponding hydrogen loaded clusters are stable and correspond to minima on the potential energy surface. Each of the Li center can adsorb three hydrogen molecules. The analysis of studied thermodynamic parameters reveals that the hydrogen adsorption on the $(N_4C_3H)_6Li_6$ cluster is favorable within a given temperature pressure zone as highlighted in the associated (T-P) phase diagram. Therefore $(N_4C_3H)_6Li_6$ cluster can be used as a potential hydrogen trap. The temperature and pressure have prominent effect on the spontaneity of the adsorption process. The 3D moiety can also be executed as a potential hydrogen trapping material.

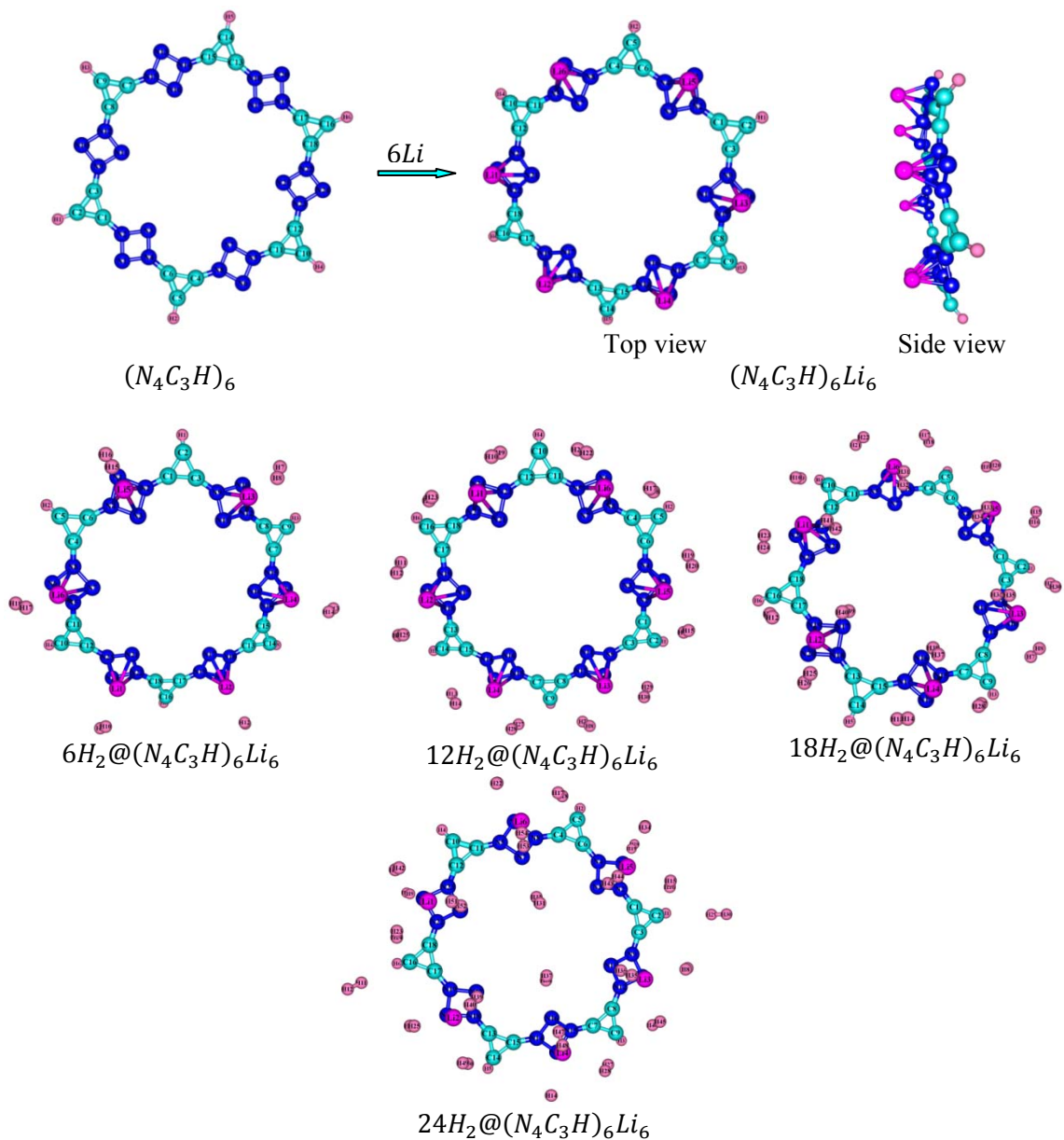

**Figure 1.** Optimized geometries (B3LYP/6-31G (d)) of $(N_4C_3H)_6$, $(N_4C_3H)_6Li_6$ and $nH_2@(N_4C_3H)_6Li_6$ $(n = 6, 12, 18, 24)$ clusters.

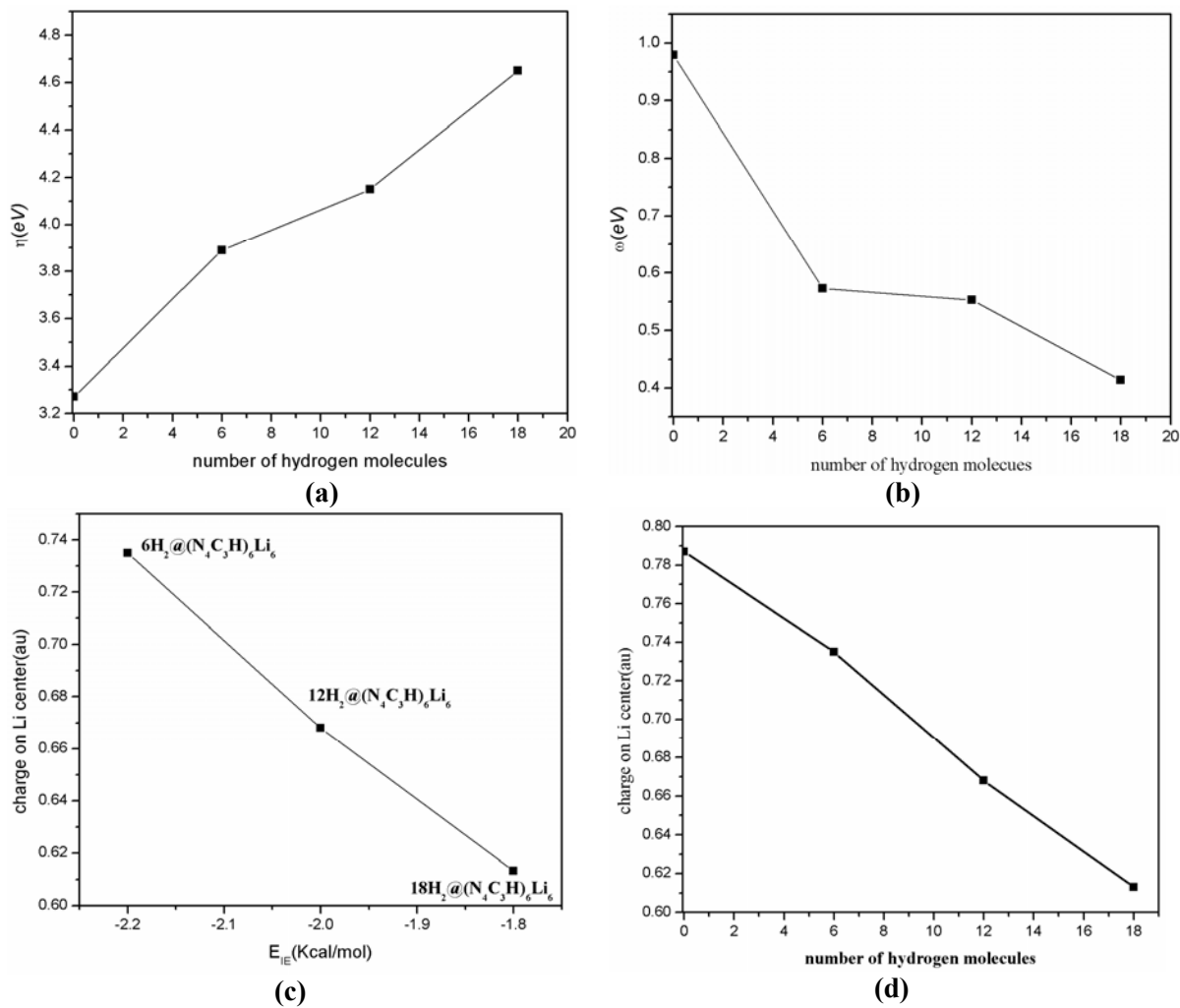

**Figure 2.** Plot of (a) hardness (η, *eV*) vs number of hydrogen molecules (b) electrophilicity (ω, *eV*) vs number of hydrogen molecules (c) charge on Li center (au) vs interaction energy per hydrogen molecule ($E_{IE}$, Kcal/mol) (d) charge on Li center (au) vs number of hydrogen molecules.

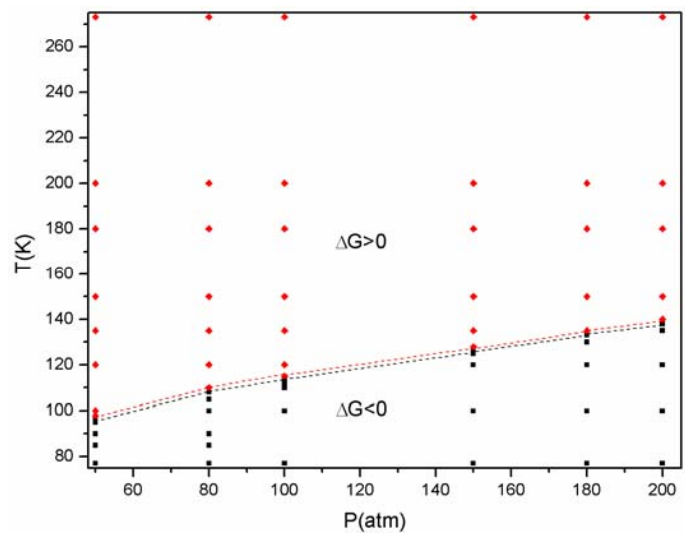

**Figure 3. Temperature–Pressure phase diagram showing the variation of ΔG.**

| Top view | 3D view |
|---|---|

(a)

| Top view | 3D view |
|---|---|

(b)

**Figure 4.Optimized geometry of (a) 3D functional material and (b) corresponding hydrogen bound analogue (with NIMAG=1).**

Table 1:  Point group (PG), total energy (E, au), electronegativity ($\chi$, $eV$), hardness ($\eta$, $eV$) electrophilicity ($\omega$, $eV$) values for $(N_4C_3H)_6Li_6$ and $nH_2@(N_4C_3H)_6Li_6$ ($n = 6, 12, 18$) clusters.

| SYSTEM | PG | E(au) | $\chi(eV)$ | $\eta(eV)$ | $\omega(eV)$ |
|---|---|---|---|---|---|
| $(N_4C_3H)_6Li_6$ | $C_{6v}$ | -2047.10014 | 2.530 | 3.271 | 0.979 |
| $6H_2@(N_4C_3H)_6Li_6$ | $C_{6v}$ | -2054.17361 | 2.111 | 3.889 | 0.573 |
| $12H_2@(N_4C_3H)_6Li_6$ | $C_1$ | -2061.24325 | 2.142 | 4.150 | 0.553 |
| $18H_2@(N_4C_3H)_6Li_6$ | $C_1$ | -2068.31032 | 1.961 | 4.646 | 0.414 |

Table 2: Interaction energy per hydrogen molecule ($E_{int}$, Kcal/mol), average chemisorptions energy ($E_{CE}$, Kcal/mol) and reaction enthalpy ($\Delta H$, Kcal/mol) values for $(N_4C_3H)_6Li_6$ and $nH_2@(N_4C_3H)_6Li_6$ ($n = 6, 12, 18$) clusters.

| SYSTEM | $E_{int}$(Kcal/mol) | $E_{CE}$(Kcal/mol) | $\Delta H$(Kcal/mol) |
|---|---|---|---|
| $6H_2@(N_4C_3H)_6Li_6$ | -2.2 | 2.2 | -5.0 |
| $12H_2@(N_4C_3H)_6Li_6$ | -2.0 | 2.0 | -7.7 |
| $18H_2@(N_4C_3H)_6Li_6$ | -1.8 | 1.8 | -9.2 |

Table 3: Charge on the lithium center and NICS (0) values of $(N_4C_3H)_6Li_6$ and $nH_2@(N_4C_3H)_6Li_6$ ($n = 6, 12, 18$) clusters.

| SYSTEM | Charge on Li center(au) | NICS(0) (ppm) |
|---|---|---|
| $(N_4C_3H)_6Li_6$ | 0.787 | 0.29 |
| $6H_2@(N_4C_3H)_6Li_6$ | 0.735 | 0.23 |
| $12H_2@(N_4C_3H)_6Li_6$ | 0.668 | 0.25 |
| $18H_2@(N_4C_3H)_6Li_6$ | 0.613 | 0.32 |